\documentclass[aps,prx,twocolumn]{revtex4-2}

\usepackage{amsmath,amssymb,graphicx,color,bm,epstopdf}
\usepackage{bbold}
\usepackage{braket}
\usepackage{enumitem}
\usepackage{orcidlink}
\usepackage{tikz}
\usetikzlibrary{shapes,arrows,decorations.pathmorphing}
\usepackage{dsfont}
\usepackage{stmaryrd}
\usepackage{mathtools}

\newlist{UR}{enumerate}{1}
\setlist[UR]{label=[M\arabic*]}
\usepackage[resetlabels,labeled]{multibib}
\newcites{S}{Supplementary References}

\begin{document}

\title{Higher order weak values for paths in nested Mach-Zender interferometers}
\author{Shushmi Chowdhury\orcidlink{0009-0005-7658-713X}}
\author{Jörg B.~Götte\orcidlink{0000-0003-1876-3615}}
\affiliation{%
 School of Physics and Astronomy, University of Glasgow, Glasgow, G12 8QQ, United Kingdom}%

\begin{abstract}
We consider weak values in the Feynman propagator framework, to gain a broader understanding of their interpretation in terms of path integrals. In particular, we examine the phenomenon of seemingly discontinuous paths that particles take in  nested Mach-Zender interferometer experiments. We extend on existing path integral approaches for weak values by deriving expressions to model a sequence of weak measurements, and study the probe shifts across the different branches of a weak value interferometer. We apply this to scrutinise two scenarios of interest, one which treats photons as measurement apparatus via their spatial projection operators, and the second treating mirrors as probes. 
\end{abstract}

\maketitle

\section{Introduction}

In standard quantum mechanics, the act of ``measurement" collapses a quantum state, initially in some state of superposition, into only one of the possible eigenstates of some given observable. Subtleties around unsettled questions, such as why we only have access to probabilistic quantities when physical systems and measurement devices seem to have deterministic wave functions, have encircled the quantum \emph{measurement problem}, leading to several competing interpretations of quantum mechanics. The notion of a \emph{weak} measurement, initially proposed by Aharonov, Albert and Vaidman (ABL) \cite{ABL}, extends the standard concept of a measurement to include weak values that result from considering the average of a quantum system's pre-selected and post-selected states when it couples weakly to measurement apparatus. 

Weak measurements enable an observer to deduce wave function properties with minimal disturbance, allowing us in a sense to circumvent the complete collapse of the wave-function in specific set-ups. They can be harnessed to probe quantum systems as they evolve in between measurements, through precise tuning of pre-selection and post-selection, and amplify otherwise undetectably small signals \cite{sup19, wv21, wm21}. Thus, weak measurements have seen potential precision measurement applications in areas ranging from sensitive high-resolution NMR spectroscopy of nuclear spins to sub-Hertz resonance \cite{wv18, wmNat19, qu2020sub}. In the context of optical beam shifts, classical analogues of weak measurements can be demonstrated in paraxial beams, where the initially homogenous polarization of a beam becomes weakly inhomogenous on reflection \cite{goos47, jorg12, goos13, wm-goos20}. Furthermore, for qubits in realistic environments in a quantum information processing context, incorporating weak measurements in schemes for fully characterising many-body correlations in a bath to arbitrary order could allow for optimising quantum control against decoherence \cite{wm04, praQC, coto2017power, prlQC, wmQC20, prxQC, wmQC24}. 

Path integrals, ever since their inception in the early 1940s resulting from early work by Richard Feynman \cite{Fey48}, have stood out as a particularly intuitive formalism for describing quantum mechanics, with its unique ability to account for all the possible trajectories (`paths') that can be taken by a particle between its initial and final states. Exploring weak values within the path integral framework has the potential of offering an elegant framework to study quantum states in superposition while taking into account all the possible paths for the transition amplitude for each state without completely collapsing the given system via strong measurement. 

We investigate the possible families of paths that particles traverse within the different branches of a nested Mach-Zehnder interferometer. Our study encompasses two distinct scenarios: the first examines photons as measurement devices via their projection operators, and the second explores mirrors as probes. To that end, we extend Matzkin's path integral \cite{matz1, Weak-Measurements-Semiclassical-Regime} approach for weak values to a sequence of weak measurements and study the probe shifts across the different branches of the interferometer. We also further consider the discontinuous photon trajectories as reported by Danan et al \cite{Asking-Photons-Where-They-Have-Been}, a result which has sparked curious discussions.


\section{Path Integral Formalism}
\label{setup}
\begin{figure*}
\includegraphics[scale=0.3]{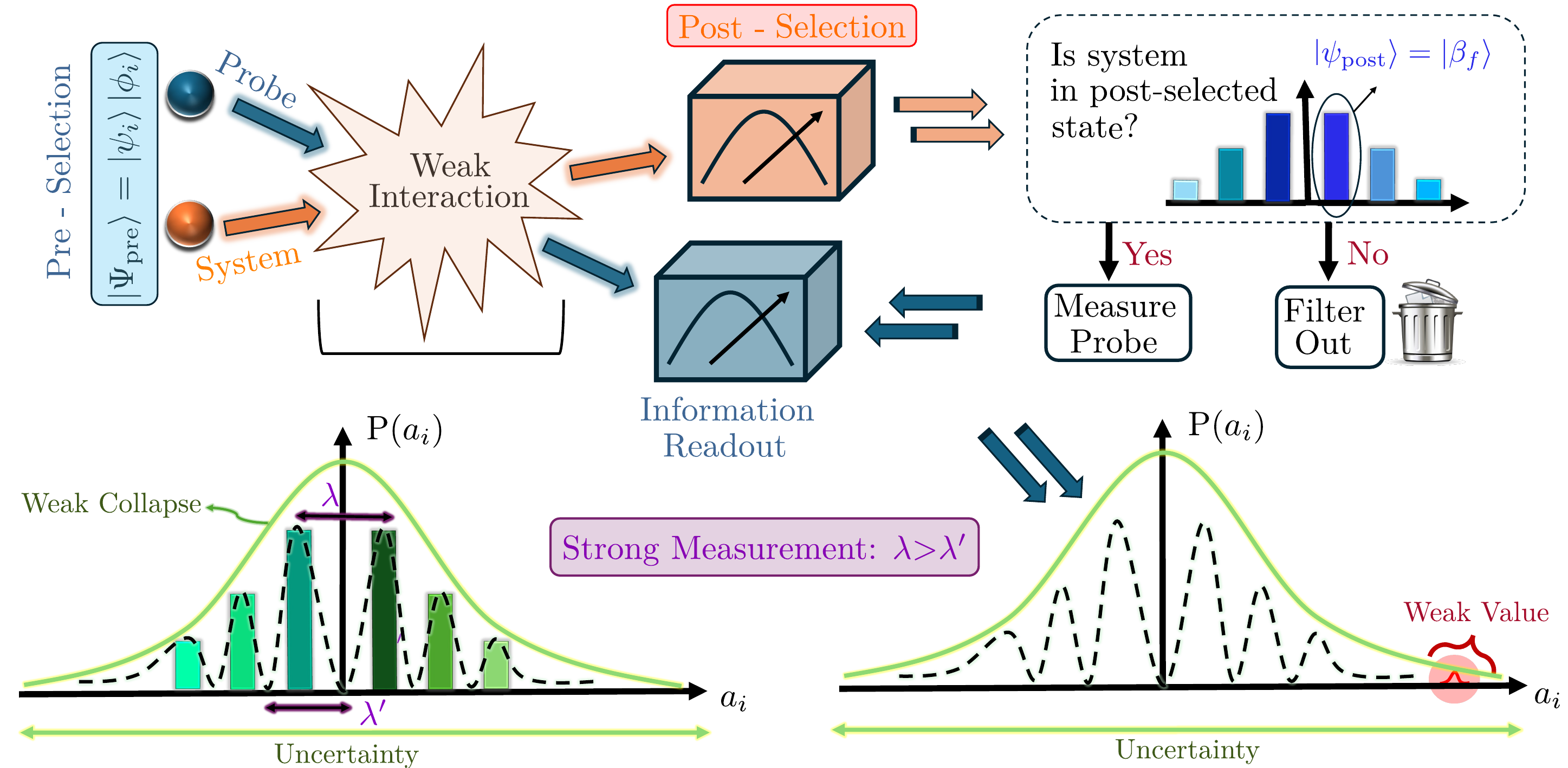}
\caption{A schematic outline of a weak measurement set-up, with the system and probe initially evolving independently, until they weakly interact. Then they couple, with the interaction causing the discrete set of eigenvalues to collapse to approximate measurement outcomes, as shown by the blue graph. Weak values arise when the system is contingent upon pre-selection and post-selection (on a different observable of the system), and the probability of obtaining this value is very small.}
\label{fig:weak}
\end{figure*}

\begin{figure} [b]
    \centering
    \includegraphics[scale=0.37]{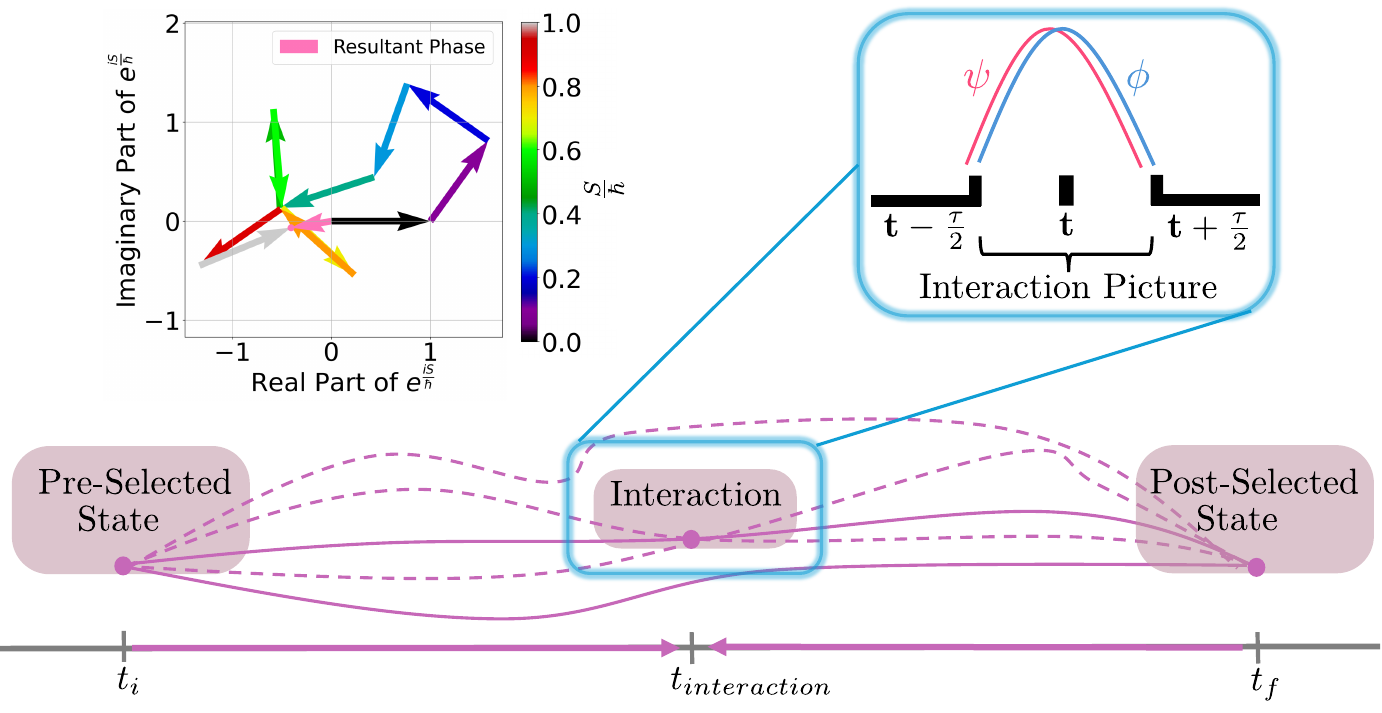}
    \vspace{-0.5em}
    \caption{A schematic outlining paths traversing from the pre-selected state to the interaction region, where system and probe wavefunctions overlap in the interaction picture, and then propagate back to the interaction region from the post-selected state. The solid lines show the classical paths and the dotted lines show the quantum paths. A figure illustrating the phase oscillations is shown on the top left, with phase-dependent path-contributions. With increasing action, $S$, only the classical path remains.}
    \label{interaction picture}
    \vspace{-1em}
\end{figure}

In Fig.~(\ref{fig:weak}), we outline a general weak measurement process. Weak values resulting from these measurements are a consequence of the system being in a pre-selected and post-selected ensemble. The quantum nature of the system allows it to take different possible trajectories simultaneously, thus resulting in a probability amplitude which is the sum of all the possible paths a system can take with each path having some phase factor associated with it.
The probability amplitude is the aggregate of individual amplitudes associated with each conceivable path, with the individual phases dictating the contribution of their respective paths to the overall amplitude. The total probability amplitude of a system to go from a point in spacetime, $x(t_A)$ to a point $x(t_B)$ is given by
 \begin{equation*}
     K(B,A) = \sum_{x(t_A)}^{x(t_B)} \text{Amplitude}[x(t)].
\end{equation*}
 where the amplitude of each individual path is $\text{Amplitude}[x(t)] = G\exp{\left\{\frac{i}{\hbar}S\left[x(t)\right]\right\}}$ with $G$ being a constant needed for normalisation. 
 The phase determining the contribution of each path is proportional to the action, $S[x(t)]$. 
 which is the integral of the Lagrangian over time in the configuration space  and is defined as $S = \int_{t_a}^{t_b} L(\dot{x},x,t) dt.$ The probability to go from $x(t_A)$ to $x(t_B)$, is then simply the square of the total amplitude,
 \begin{equation*}
      P(B \leftarrow A) = |K(B,A)|^2 
 \end{equation*}
When the action in the phase is much greater than Plank's constant, $S\gg \hbar$,  the `quantum' paths oscillate rapidly, destructively interfering. Thus,  in this limit,  the classical path remains, as shown top left in Fig.~(\ref{interaction picture}). 

Similar to the formalism introduced by Matzkin et. al. \cite{matz1}, we consider a separable wavefunction, 
\begin{equation}
\ket{\Psi(t_i)} = \ket{\psi_i}\ket{\phi_i},
\end{equation}
combining the initial states of the system and the probe, with $\ket{\psi_i}$ and $\ket{\phi_i}$ denoting the pre-selected states of each respectively. With the initial states assumed to be Gaussian, the system and probe are weakly coupled through a weak von Neumann interaction Hamiltonian \cite{vonNeumann1955}, 
\begin{equation}
    H_\text{interaction} = g(t) {A} {P}f_\text{interaction}, 
\end{equation}
where $g(t)$ represents a function reflecting the interaction duration, $\tau$, centred on the interaction time, $P$ and $A$ are the probe momentum and observable respectively, and $f_\text{interaction}$ is a Dirac delta function that is only non-vanishing near the interaction region. We denote the interaction time $t_{interaction}$ as $\textbf{t}$, which gives us
\begin{equation}
g(t)=\int_{\textbf{t}-\tau/2}^{\textbf{t}+\tau/2} g(t') \,dt' , 
\end{equation}
with the system and the probe evolving independently until they reach the interaction region, where they then weakly couple.
In the path integral framework, each path is weighted by the potential of the system at the interaction point. The total Lagrangian describing the setup from pre-selection to post-selection is given by
\begin{equation*}
    L_\text{total}=  \underbrace{\frac{\dot{x}^2 m}{2} - V(\textbf{x})}_{L_\text{system}} + \underbrace{\frac{\dot{X}^2 M}{2}}_{L_\text{probe}} - \underbrace{g(t){A}(x)M\dot{X}f(x,X)}_{L_\text{interaction}}.
\end{equation*}
In the Schrödinger picture, the state vectors of both the system and the probe evolve freely until the interaction time \textbf{t}, and then continue to the final time. The interaction, which occurs over a time interval, $\tau$, introduces a time dependency, acting as a small perturbation in the Schrödinger Eq. The total evolution operator is given by the system from the initial to the final time
\begin{equation}
\begin{aligned}
U_{\text{total}} &= U_{\text{free}} \left( t_f, t + \frac{\tau}{2} \right) U_{\text{interaction}} \left( t + \frac{\tau}{2}, t - \frac{\tau}{2} \right) \\
&\hspace{5em}\quad U_{\text{free}} \left( t - \frac{\tau}{2}, t_i \right)
\end{aligned}
\end{equation}
where $U_{total}$ evolves the system from the initial time, $t_i$, to the final time, $t_f$. The total Hamiltonian is  
\begin{equation}H_{\text{total}} = H_{\text{system}} + H_{\text{probe}} + H_{\text{interaction}},
\end{equation}
with \( H_{\text{interaction}} \) treated as a perturbation to the free evolution 
($H_0 = H_{\text{system}} + H_{\text{probe}}$), in the interaction picture, see Fig~[\ref{interaction picture}]. In position space, the evolution of a quantum state over time is described by the propagator relation
\begin{equation*}
    \psi(x,t) = \int K(x,t; x_i,t_i) \psi(x_i,t_i) dx_i.
\end{equation*}
The system and the probe weakly couple upon interaction, resulting in a non-separable propagator that encompasses the full transition,
\begin{equation}
\begin{aligned}
K_{total}\Big( X_f, x_f, t_f; X_i, x_i, t_i \Big) =  \\
&\hspace{-8em}\quad \quad \int D[x(t)]D[X(t)]  
e^{\frac{i}{\hbar} \int_{t_i}^{t_f} L_{total}}
\label{k_total}
  \end{aligned}  
\end{equation}
Given that the interaction is weak, the interaction Lagrangian in Eq.~(\ref{k_total}) can be expanded to first order,
\begin{align}
K_{total} &= \int D[x(t)]D[X(t)] \exp{ \left[ \frac{i}{\hbar} \int_{t_i}^{t_f}  L_{system} + L_{probe} \right]} \nonumber \\
&\hspace{1.5em} \left[{1-\frac{i}{\hbar}\int_{t_i}^{t_f}g(t){A}(x)M\dot{X}f(x,X)}\right].
\end{align}

\begin{figure*}    
\centering 
    \includegraphics[width=\textwidth]{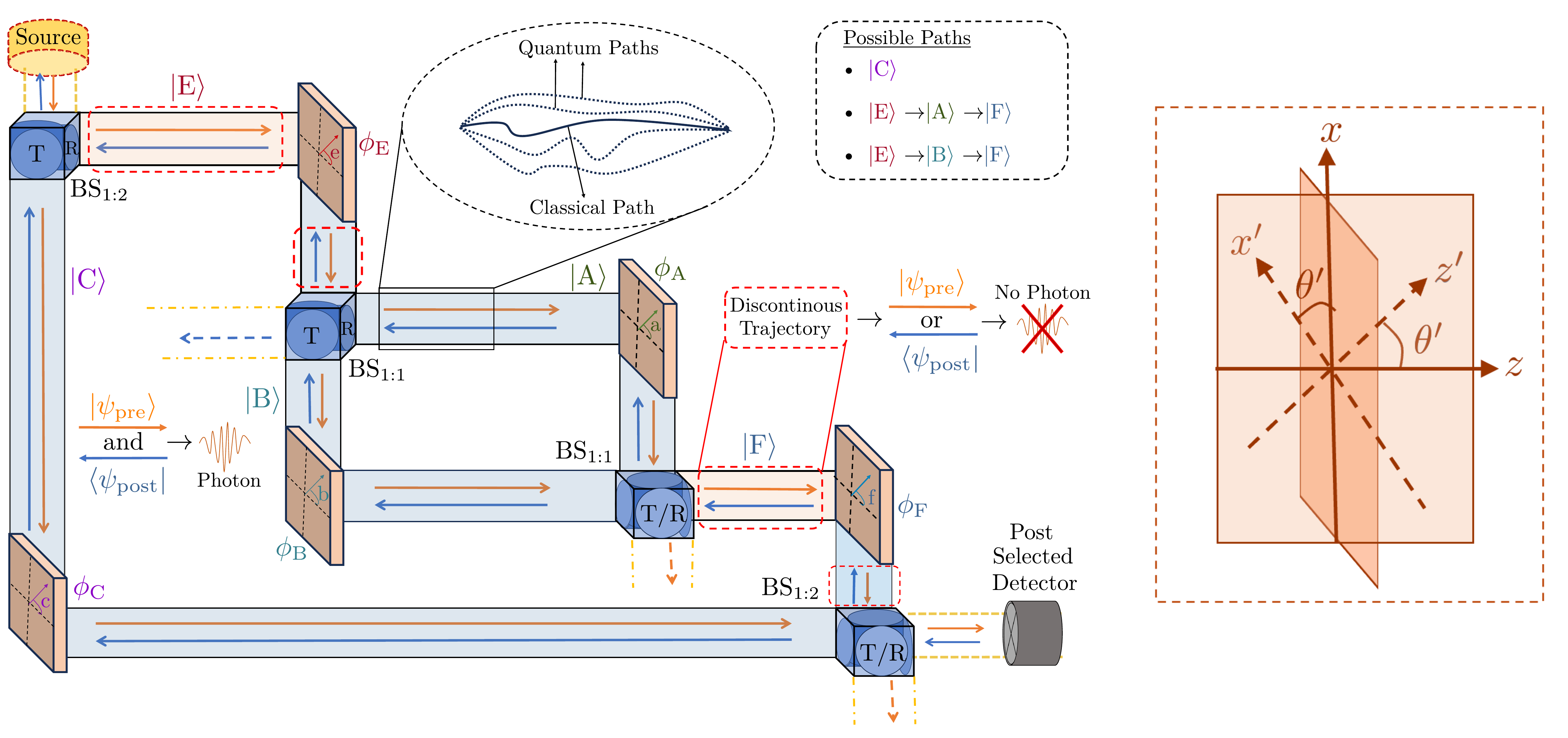}
    \caption{The schematic depicts all potential photon paths within a nested interferometer. Each possible path is encompassed within its own family of paths. Orange arrows show a pre-selected state propagating forward from the source and blue arrows show a backward propagating wave from the post-selected detector.}
    \label{fig:nested-interferometer}
\vspace{-1.5em}
\end{figure*}

The pre-selected state, $\ket{\Psi(x_\text{i},t_\text{i})}$, is propagated to interaction time, 
\begin{equation}
U(\textbf{t},t_i) \ket{\psi(x_\text{i}, t_\text{i})} \ket{\phi(x_\text{i}, t_\text{i})} = \ket{\psi(\textbf{x}, \textbf{t})}\ket{\phi(\textbf{X}, \textbf{t})},
\end{equation}
while the post-selected state, $\ket{\Psi_f(x_\text{f},t_\text{f})}$, is propagated backward to interaction time,  
\begin{equation}
\bra{\psi(x_\text{f}, t_\text{f})} \bra{\phi((x_\text{f}, t_\text{f})} U^*(t_f,\textbf{t}) = \bra{\psi(\textbf{x}, \textbf{t})}\bra{\phi(\textbf{X}, \textbf{t})}.
\end{equation}
At the interaction region, the weak coupling between the system and the probe results in a weak measurement of observable $A$  on the system. At the final time, a strong measurement of observable $B$ is made, with only the selected outcomes $\ket{\beta}$ retained for post-selection. Here the state vectors are propagated state of the wave function (for detail see supplement of \cite{matz1}).

The probe reveals information about the system obtained during the weak interaction through a strong projection onto the final probe state, $\ket{X_f}$.  The probe state, $\ket{X_f}$, is contingent on the post-selected state, $\ket{\beta}$, with a final probe state given by
\begin{equation}
\bra{X_f}\bra{\beta}U^\dagger(t_f,\textbf{t})U(\textbf{t}, t_i)\ket{\psi_i}\ket{\phi_i} = \phi_\beta (X_f,t_f).
\end{equation}
The weak value is contained in the probability amplitude, $\phi_{\beta}(X_f,t_f)$. The probe shift due to the interaction is proportional to the real part of the weak value \cite{col}, with the interaction between the system and probe resulting in the final state being a shifted Gaussian, 
\begin{equation}
\ket{\phi_f}=e^{-igA^w\hat{P}}\ket{\phi_i}.
\end{equation}
After interaction with the photon, the probe evolves, weakly coupled to the system up to the final detection time. The probe shift takes place at the interaction time, $\textbf{t}$, on interaction with the system and is evaluated \cite{matz1} at the final time, $t_f$, as,
\begin{equation}
\begin{aligned}
\phi_{\beta_f}(X_f,t_f) &= \int_{\textbf{X}}^{X_f} D\big[Q(t)\big] \exp\Bigg\{\frac{i}{\hbar} \int_{\textbf{t}}^{t_f} L dt' \Bigg\} \\
&\hspace{-5em}
\times \phi(\textbf{X},\textbf{t}) \times \int d\textbf{x} dx_f K_s^*\Big(x_f,t_f;\textbf{x},\textbf{t}\Big) b_f^*(\textbf{x},\textbf{t}) \times \psi(\textbf{x},\textbf{t})\\
&\hspace{-5em}\times \exp\Bigg\{- \frac{ig}{\hbar} M\dot{Q} 
\\
&\hspace{-5em}\underbrace{\Bigg(\frac{\int d\textbf{x} dx_f K_s^*\Big(x_f,t_f;
\textbf{x},\textbf{t}\Big) \beta_f^*(\textbf{x},\textbf{t}) A(q)f(q,X)\psi(\textbf{x},t)}{\int d\textbf{x} dx_f K_s^*\Big(x_f,t_f;\textbf{x},\textbf{t}\Big) \beta_f^*(\textbf{x},\textbf{t})\psi(\textbf{x},\textbf{t})}\Bigg)}_{  \textbf{A}^\textbf{w}   }\Bigg\} 
\end{aligned}
\end{equation}

To include multiple interactions with the system and probe we slightly alter the separable wavefunction as
\begin{equation}
\ket{\psi_i} \bigotimes_{n= A,B,C \dots} \ket{\phi_{(n)i}}
\end{equation}
where the alphabets represent the interaction regions, which in our case correspond  mirrors inside a Mach-Zender Interferometer.

\section{Particle Paths in a Nested Mach-Zender Interferometer}
\label{Particle Paths in a Nested Mach-Zender Interferometer}
We now consider the discontinuous paths traversed by particles within the Mach-Zehnder Interferometer, as studied previously by Vaidman \cite{Past-of-a-quantum-particle} using weak traces, followed by a re-evaluation of subsequent experimental evidence \cite{Asking-Photons-Where-They-Have-Been}. In Fig.~(\ref{fig:nested-interferometer}), we show a nested Mach-Zehnder Interferometer where each of the mirrors is tilted at specific angles. These angles can be considered equivalent to oscillating mirrors with distinct frequencies, as used in \cite{Asking-Photons-Where-They-Have-Been}. On hitting the mirrors, the photons acquire shifts that are Fourier-transformed to the frequency domain in the post-selected quad-cell detector, allowing to track the photons hitting the different mirrors simultaneously  \cite{Asking-Photons-Where-They-Have-Been, Past-of-a-quantum-particle, Tracing-the-past-quantum-particle}. The spectra reported with the setup can be found in \cite{Asking-Photons-Where-They-Have-Been, Tracing-the-past-quantum-particle}, resulting from different photons traversing in different branches.

Mirror B is tuned to get destructive interference at the second beam splitter in the inner interferometer, preventing the beam from reaching mirror F and thus, the detector. One would naturally assume that the photon traveled through the lower path of the interferometer if detected by the post-selected detector, since it seems rational that only the signal from mirror C would trigger it. However, the spectrum in \cite{Asking-Photons-Where-They-Have-Been,  Tracing-the-past-quantum-particle}, shows that the particle was present in the inner interferometer, at mirrors A and B. The absence of the signals in mirrors $E$ and $F$ but the presence of the mirrors inside the inner interferometer led the authors to conclude that photons can follow discontinuous trajectories given that post-selection is successful. This can be explained by the Two State Vector Formalism (TSVF) \cite{Time-Symmetry}, where each photon detection is described as a post-selected quantum state $\phi$ evolving backward from the detector and an initial quantum state evolving forward from the source. The overlap of these two quantum state wavefunctions is where we find the `weak trace' of the particle. This is shown in Fig~\ref{fig:nested-interferometer} using orange arrows for the forward wave and blue arrows for the backward wave. The interactions in these overlap regions between the photon and mirrors are local, thus the weak values of the projection operators at the interaction regions will be proportional to the weak trace \cite{Past-of-a-quantum-particle}. 

Extending the weak value expression in the earlier section  to account for multiple interactions inside the Mach-Zehnder Interferometer involves considering three possible sets of interactions, with each interaction in each path being unique. There are three possible paths that a particle can take after leaving the source and before entering the post-selected detector, as follows
\begin{itemize}
\item Path 1: $\ket{\text{E}} \rightarrow \ket{\text{A}} \rightarrow \ket{\text{F}}$
\item Path 2: $\ket{\text{E}} \rightarrow \ket{\text{B}} \rightarrow \ket{\text{F}}$
\item Path 3: $\ket{\text{C}} $.
\end{itemize}

In the path integral framework, the interferometer model can be analysed by dividing it into sub-paths where the total amplitude at the detector is the sum of the amplitudes from individual paths (Path 1, Path 2, and Path 3), which then interfere with each other. Squaring the total amplitude gives the probability of detecting the photon in the detector. For simplicity, we will focus on interactions within one path and use this as a general groundwork for the other paths. As earlier, we can propagate the pre-selected state forward and the post-selected state backward to the interaction region. However, $\text{Path 1}$ contains three interaction regions, and for our analysis, we choose to propagate both pre-selected and post-selected states to mirror A. The same process could be applied to mirror E or B with identical outcomes, as illustrated in Fig.~(\ref{fig:sim_mir}) in the supplemental material. Similarly as in \cite{matz1}, we characterise the interactions by
\begin{equation*}
L_\text{int}{\phi_E} = g(t)A(x)f(x,X_{E})M_E\Dot{X_E} = \epsilon_E\\
\end{equation*}
\vspace{-1em}
\begin{equation*}
L_\text{int}{\phi_A} = g(t)A(x)f(x,X_{A})M_A\Dot{X_A} = \epsilon_A\\
\end{equation*}
\vspace{-1em}
\begin{equation*}
L_\text{int}{\phi_F} = g(t)A(x)f(x,X_{F})M_F\Dot{X_F} = \epsilon_F\\
\end{equation*}
where the coupling, $g(t)$, is the effective coupling. The total Lagrangian for this path is thus given by
\begin{equation}
    L_\text{tot} = L_E +L_A+L_F-L_{\text{int}-E}-L_{\text{int}-A}-L_{\text{int}-F},
\end{equation}
and considering weak Von Neumann interactions, we take the first order expansion, giving
\begin{equation}
\begin{aligned}
\exp{\left(\frac{i\int L_\text{tot} dt}{\hbar}\right)} =   \\
&\hspace{-8em}\quad \quad \int \exp{\left(L_E +L_A+L_F\right)}
(1-\epsilon_E)(1-\epsilon_A)(1-\epsilon_F).
\end{aligned}  
\end{equation}
The quantum state of the wavefunction after the interaction with the last mirror in this sequence is given by (see \cite{Weak-Measurements-Semiclassical-Regime} and the supplementary material for derivation),
\begin{align*}
\Ket{\Psi(t_3 + \tau/2)} &= U(t_3, t_2) \exp(-i \epsilon_3) U(t_3, t_2) \exp(-i \epsilon_2)  \\ 
&\quad \hspace{-7em} U(t_2, t_1) \exp(-i \epsilon_1) U(t_1, t_i) \Ket{\Psi(t_i)} \Ket{\phi_E(t_i)} \Ket{\phi_A(t_i)} \Ket{\phi_F(t_i)}.
\end{align*}

Drawing from the detailed calculations presented in section~\ref{WS} in the supplementary material, the weak value as the cumulative sum of the individual weak values generated at each interaction point with the mirrors is
\begin{align}
A^w_{path 1} &= A^w_{\phi_E} + A^w_{\phi_A} + A^w_{\phi_F} \nonumber \\
& \hspace{-2em}=\frac{\int K_s(x_f, t_f; x_i, t_i)_E b^*_f(t_f, t_f) [\boldsymbol{\epsilon}_{E}] \psi(x_i, t_i)}{M_E\dot{X}_E g\int K_s(x_f, t_f; x_i, t_i) b^*_f(x_f, t_f) \psi(x_i, t_i)} \nonumber \\
&\quad \hspace{-2em}+ \frac{\int K_s(x_f, t_f; x_i, t_i)_A b^*_f(x_f, t_f) [\boldsymbol{\epsilon}_{A}] \psi(x_i, t_i)}{M_A\dot{X}_A g\int K(x_f, t_f; x_i, t_i) b^*_f(x_f, t_f) \psi(x_i, t_i)} \nonumber \\
&\quad \hspace{-2em}+ \frac{\int K_s(x_f, t_f; x_i, t_i)_F b^*_f(x_f, t_f) [\boldsymbol{\epsilon}_{F}] \psi(x_i, t_i)}{M_F\dot{X}_F g\int K_s(x_f, t_f; x, t_i) b^*_f(x_f, t_f) \psi(x_i, t_i)}
\label{wv}
\end{align}

\begin{figure}
\centering
\includegraphics[scale=0.35]{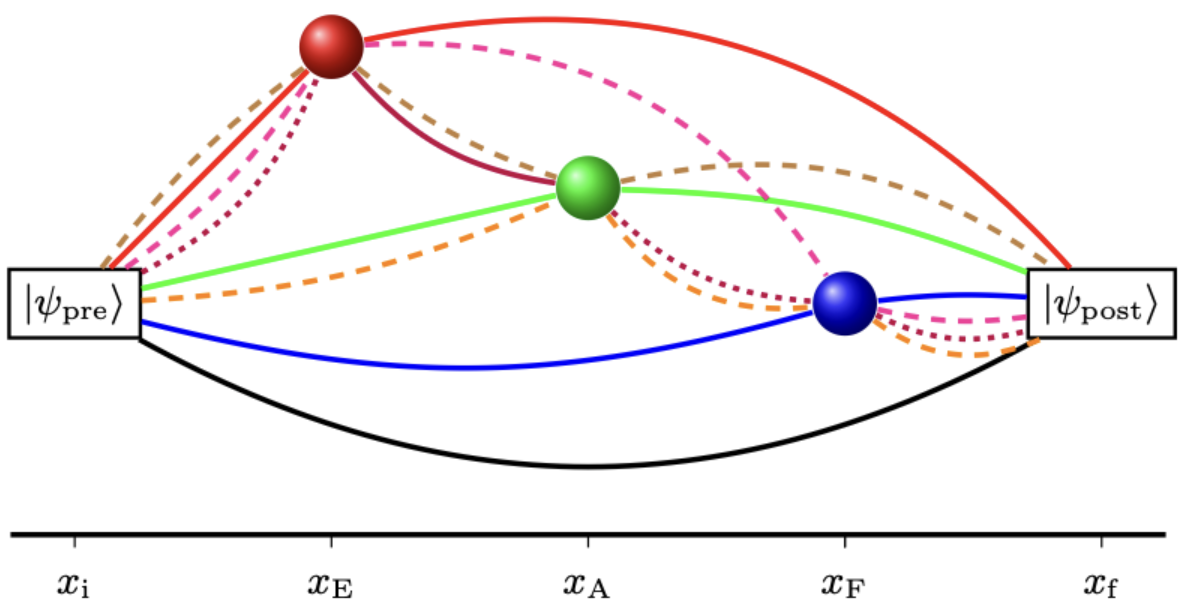}
\caption{The schematic shows the possible interaction with subsequent mirrors. The dashed lines show two subsequent interactions and the dotted line shows three subsequent interactions. The black illustrates the unperturbed path with no interaction. The first-order interactions are shown by the coloured solid lines.}
\label{nhn} 
\vspace{-1.5em}
\end{figure}

The $K(x_f;x_t)_n$ terms in the numerator are the propagators passing through the mirror regions with $n =A, B, C, E, F$. For example, 
$K(x_f;x_i)_A = K(x_f;x_A) K(x_A;x_i)$, for detailed derivation the reader is referred to section~\ref{WS}. The $K(x_f;x_t)$ in the denominator sums up all the paths that do not interact with any mirrors. The $b^*$ is the post-selected state propagated backwards. Because epsilon is very small, scaling amplitudes by  $\epsilon^2$ and $\epsilon^3$ results in even smaller probabilities of orders $\epsilon^4$ and $\epsilon^6$ respectively. Given the minimal contributions of these paths to observable outcomes, it is justified to focus on the first-order terms of Eq.~(\ref{wv}). Including the shifts from the other paths mentioned previously, $\ket{\text{C}}$ and $\ket{\text{E}} \rightarrow \ket{\text{B}} \rightarrow \ket{\text{F}}$, we determine the total cumulative shift to first order arising from these subsequent weak interactions in the post-selected detector,
\begin{align}
A^w_{\text{detector}} &= A^w_{\phi_A} + A^w_{\phi_B} + A^w_{\phi_C} \nonumber \\
&\quad \hspace{-4em}+ \underbrace{\frac{\int K_s(x_f, t_f; x_i, t_i)_E b^*_f(x_f, t_f)\left[(\boldsymbol{\epsilon+\epsilon})_E\right] \psi(x_i, t_i)}{M_E\dot{X}_E g\int K_s(x_f, t_f; x_i, t_i) b^*_f(x_f, t_f) \psi(x_i, t_i)}}_{A^w_{\phi_E}} \nonumber \\
&\quad  \hspace{-4em}+\underbrace{\frac{\int K_s(x_f, t_f; x_i, t_i)_F b^*_f(x_f, t_f) \left[(\boldsymbol{\epsilon+\epsilon})_F\right] \psi(x_i, t_i)}{M_F\dot{X}_F g\int K_s(x_f, t_f; x_i, t_i) b^*_f(x_f, t_f) \psi(x_i, t_i)}}_{A^w_{\phi_F}}
\label{90}
\end{align}

The spectrum reported in experiment (see Fig.~2(a) in\cite{Asking-Photons-Where-They-Have-Been}) is supported by Eq.~(\ref{90}), and we get non-negative values of the projection operator of the photons at all the mirrors given that mirror B is not tuned to get destructive interference, $(P_n)_w \neq 0$, where we $n$ stands for mirrors A, B, C and so on. The weak value results from the overlap of the post-selected state determined at the final time with the pre-selected state at the interaction region. According to \cite{matz1}, the forward propagated paths for the pre-selected state in the interaction regions, E and F, are orthogonal to the backward propagated paths for the post-selected state at this region, resulting in vanishing weak values. To add to this explanation, we can see in the expression for the combined weak value at the detector, Eq.~(\ref{90}), that if we take $\epsilon=-\epsilon$ for probes E and F, the weak values disappear. In this scenario, the particle was always there on E and F, but the wavefunctions from these regions interfered in a way that made the weak trace disappear.

\begin{figure}[b]
    \centering
    \includegraphics[scale=0.25]{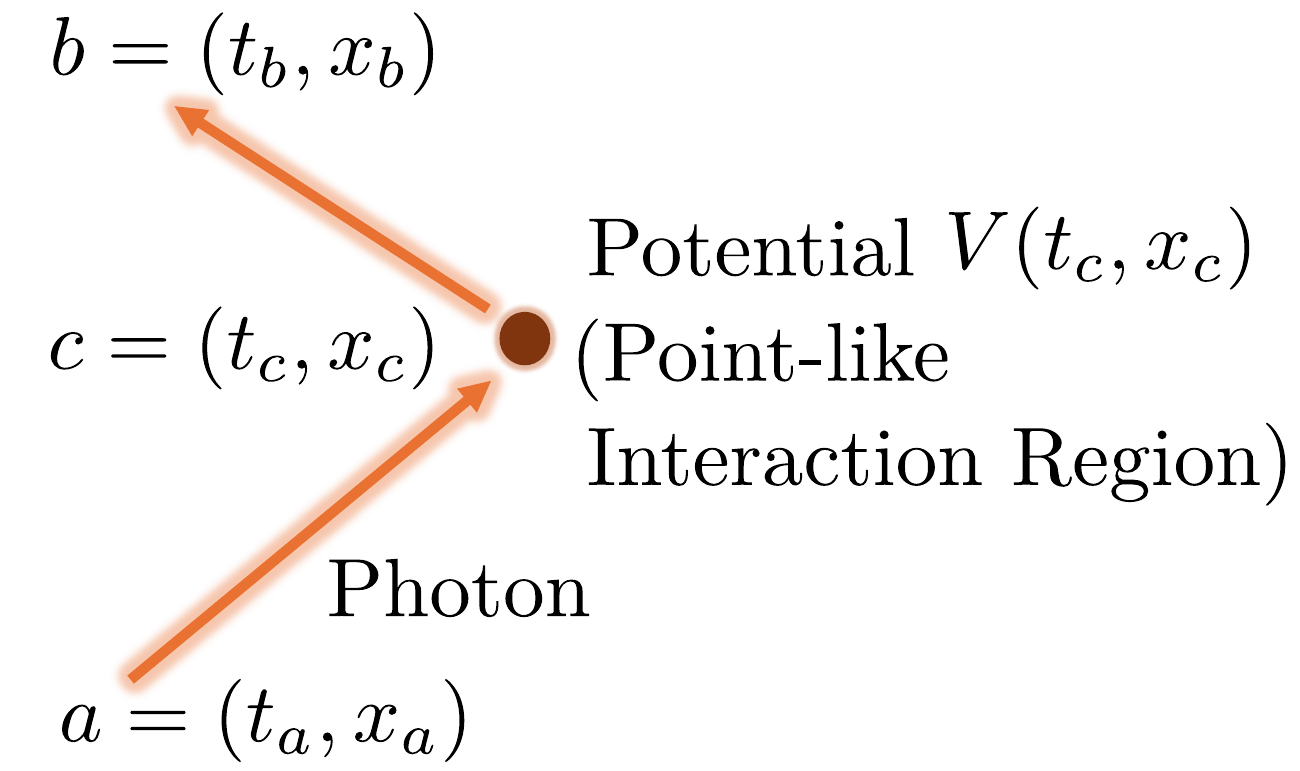}
    \vspace{-0.5em}
    \caption{A particle moves freely from $x_a$ to $x_c$, scatters at $x_c$ by potential  $V(t_c,x_c)$, then proceeds freely to $x_b$.}
    \label{path_fey}
\end{figure}

\section{Feynman Propagators and Discontinuous Trajectories}
\label{Feynman Propagators and Discontinuous Trajectories}

In the path integral framework, the interaction can be treated like a particle being scattered due to encountering a potential \cite{feynman1965flp}, as shown in Fig~(\ref{path_fey}). Paths can be constructed for successive events with amplitudes, denoted by $K(b, a)$, multiplying \cite{book-1,book-2}.

The particle trajectory can be described by \cite{feynman1965flp}
\begin{equation}
    K(b,a) = K(b,c)\left[-\frac{i}{\hbar} V(c)\right] K(c,a),
\end{equation}
and by adapting our previous methodology, we can replace the potentials with the interaction terms corresponding to each mirror and introduce a first-order expansion which gives
 \begin{align*}
    \text{Path 1'}:& \ K(x_f; x_F)(1-\epsilon_F)K(x_F; x_A)(1-\epsilon_A)\\
    & \ K(x_A; x_E)(1-\epsilon_E)K(x_E; x_i) \\
    \text{Path 2'}:& \ K(x_f; x_F)(1-\epsilon_F)K(x_F; x_B)(1-\epsilon_B) \\
    & \ K(x_B; x_E)(1-\epsilon_E) K(x_E; x_i) \\
    \text{Path 3'}:& \ K(x_f; x_C)(1-\epsilon_C)K(x_C; x_i).
\end{align*}
At the inner interferometer's second beam splitter, Path 1 and Path 2 interfere, and this recombined path further interferes with Path 3 at the outer interferometer's second beam splitter. We therefore add the interaction terms and keep track of the epsilons.
\begin{equation*}
\begin{split}
\text{Path 1'} + \text{Path 2'} + \text{Path 3'} = \\
& \hspace{-7em} 3 - \underbrace{\left(\epsilon_A + \epsilon_B +\epsilon_C \right) - \left(2\epsilon_E + 2\epsilon_F \right)}_{\text{1st order}} \\
& \hspace{-7em} +\underbrace{\left[ \left(-\epsilon_A - \epsilon_B \right)\left(-\epsilon_E - \epsilon_F\right) + 2\left(\epsilon_E\epsilon_F\right) \right]}_{\text{2nd order}} \\ 
& \hspace{-7em} +\underbrace{\left(-\epsilon_A - \epsilon_B \right)\epsilon_E\epsilon_F}_{\text{3rd order}}.
\end{split}
\end{equation*}
The zeroth term represents the summation of all the unperturbed paths. In the presence of discontinuous trajectories, the mirrors are finely adjusted within the inner interferometer such that the beam going through path 1 acquires a phase shift of $\pi$ relative to that in path 2. As a result, the wavefunctions in these two paths interfere destructively at the second beam splitter of the inner interferometer, allowing us to show this as $\text{Path 1} = -\text{Path 2}$. Building on this, first-order terms $\epsilon_E$ and $\epsilon_F$ from these paths cancel out due to destructive interference where the paths meet again. 

It is worth noting that excluding second-order terms isn't necessarily justified, so we still have contributions from $\epsilon_E$ and $\epsilon_F$ in the second order. Even if the amplitude of $\epsilon^2$ is extremely small, the probability of detecting the particle at E or F does not fall to zero insofar as second-order terms are included. It is non-trivial to associate discontinuous paths with 0 weak values without taking into account the second and higher-order terms, with work in a similar vein carried out in \cite{article1, Sokolovski}.

\section{Experimental Proposal to follow discontinuous paths}

\begin{figure}
    \centering
    \includegraphics[scale=0.30]{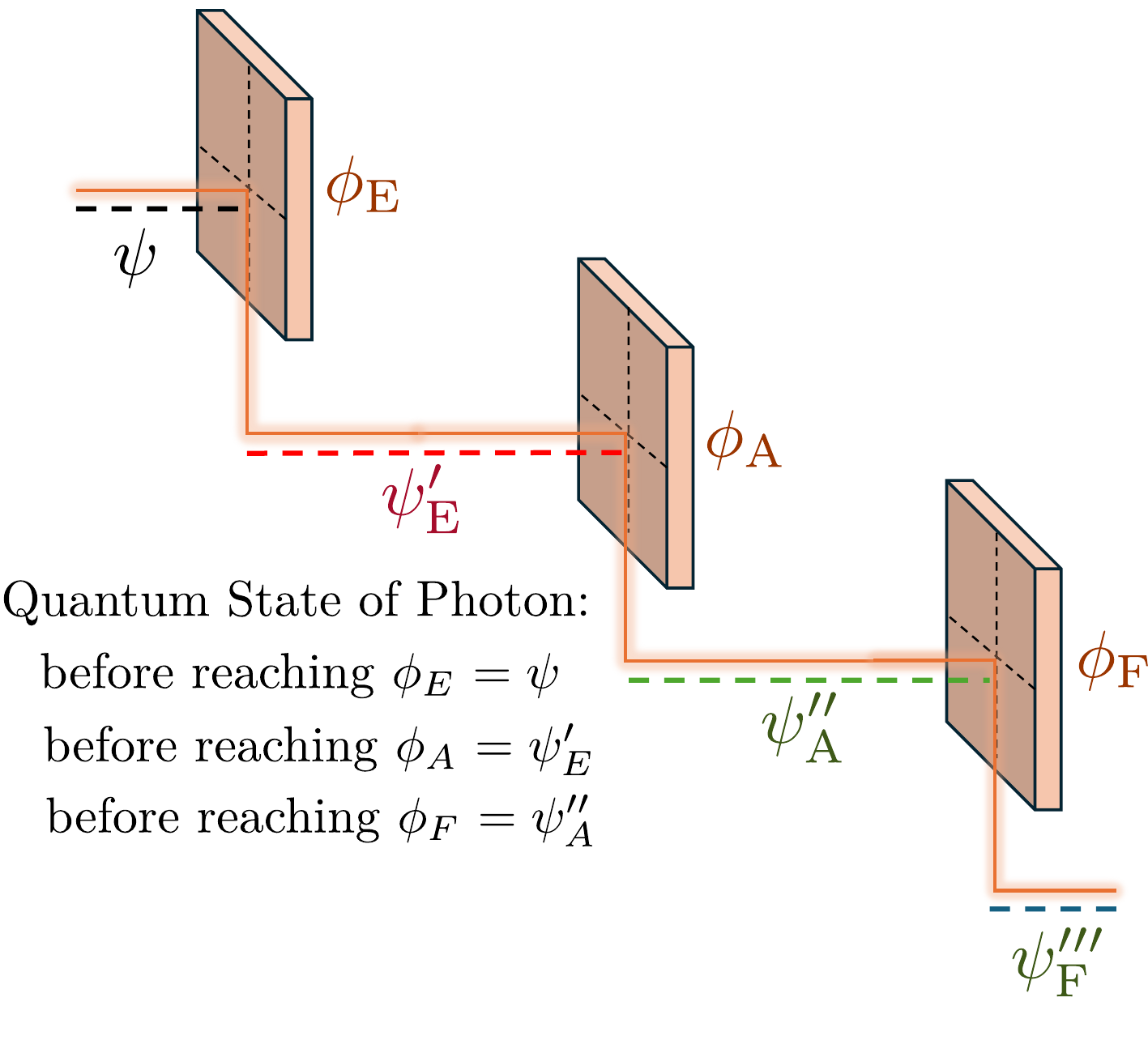}
    \vspace{-2em}
    \caption{Schematic for the quantum states of the mirror and photon change after each interaction.}
    \vspace{-1em}
    \label{see}
\end{figure}

Mirrors at different frequencies \cite{Asking-Photons-Where-They-Have-Been}, can be artificial in the sense that instead of interacting with the mirror, the interaction occurs with other degrees of freedom of the photon. The shifted Gaussian beam detected at the detector is 
\begin{equation}
    \Phi(x,z) = \exp{\left(-\frac{x^2+(z-\delta)^2}{2\sigma^2}\right)}.
\end{equation}
Experimentally, this provides efficiency as both measurement and post-selection come from the same photon. Oscillating the mirrors at different frequencies acts as a useful trick employed which lets the shifts be distinguishable at the detector. A different experiment carried out by Matkzin et al. \cite{unknown} detected shifts in spatially separated degrees of freedom in a two-arm interferometer with one weak measurement shifting the beam in a horizontal direction and the other in the vertical \cite{cheshire-cat}. It can be formally challenging to distinguish shifts from various interactions because they are of order $\epsilon$ and extremely small. On the other hand, we can focus on treating the mirrors as probes. Experimentally, this will be hard because we have to look back at the mirrors every time the post-selected detector clicks and the detector won't click most of the time owing to the nature of the post-selection.
\begin{figure*}
    \vspace{-1em}
    \centering
    \includegraphics[scale=0.35]{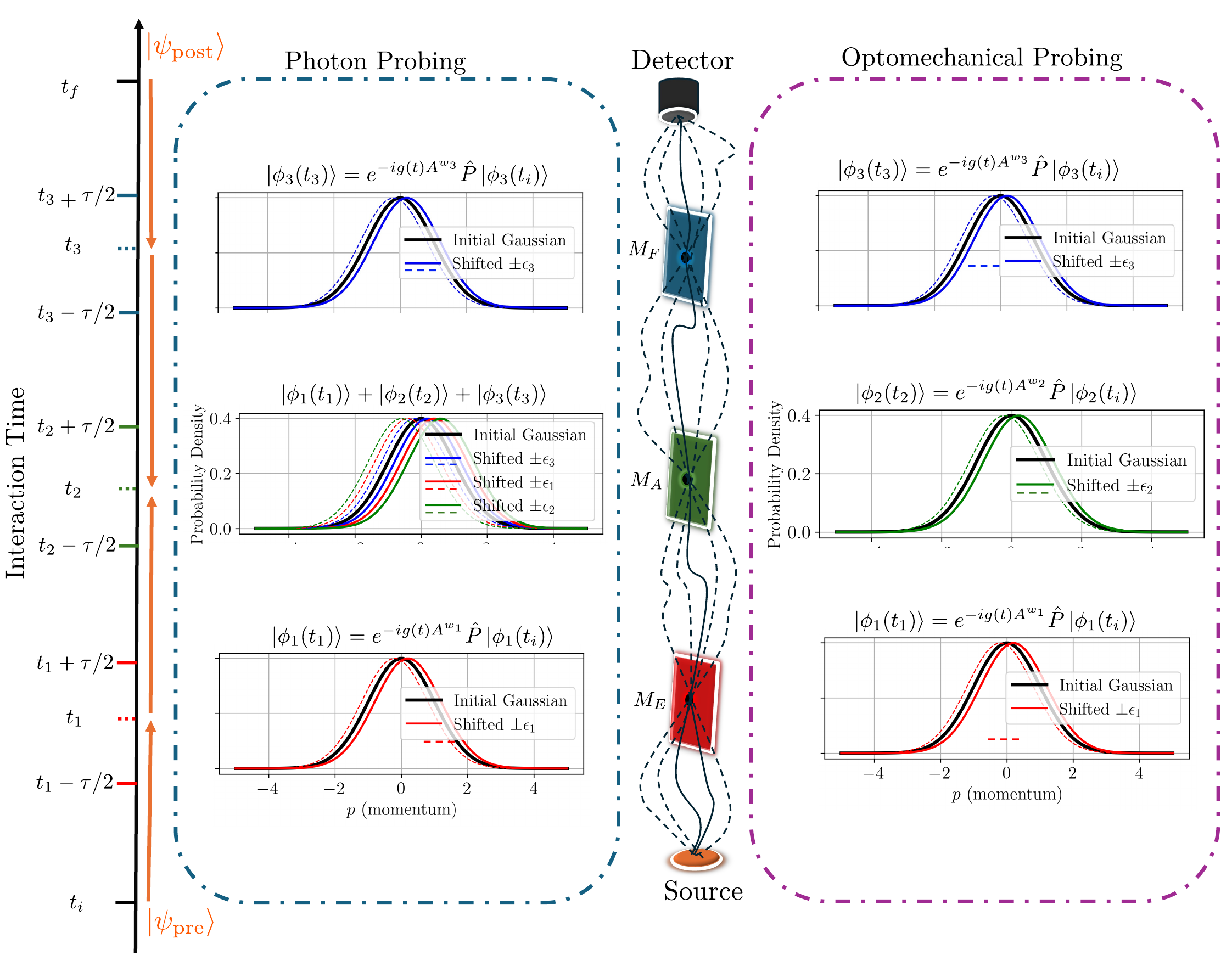}
    \vspace{-1em}
    
    \caption{On the left we show the case when the photon interacts subsequently with its spatial degree of freedom at three regions, accumulating the shifts (due to weak interactions) for the final amplitude. On the right, each mirror states acquire shifts independent of the other. In both cases, the pointers are taken to be Gaussians.}
    \label{fig:okst}
    \vspace{-1em}
\end{figure*}
For a good interferometer, the mirrors have to have a large uncertainty in momentum given by a very small uncertainty in position.  After interacting with the photon, the mirror gets a momentum kick and changes its quantum state. Using a coherent state of light, the average momentum imparted by each photon on the mirror is given by \cite{classical-limit-of-quantum-optics}
\begin{equation}
    \delta P_M = 2 \bar{n}\hbar\omega \cos(\theta')
\end{equation}
where $\bar{n}$ is the average photon number. Experimentally, we can hypothesize this momentum transfer using opto-mechanical probing. We start with the mirrors mounted on harmonic mechanical oscillators, initially on ground states, and then after the photon's interaction with the mirror, the oscillator becomes entangled with the photon. Simply put, if enough photons interact with the mirror, we will see the oscillator move from its ground state and start oscillating. An elaborate example of the above can be found in \cite{opto} where microscopic degrees of freedom are capable of making an observable effect on the macroscopic degrees. 

We can interpret the same weak value expression as the derived Eq.~(\ref{wv}) in Section~(\ref{Particle Paths in a Nested Mach-Zender Interferometer}), and consider the mirrors as probes instead of the photon's internal degree of freedom. In this case, avoiding the second-order terms means avoiding the correlation between the mirrors. The expression, in the first order, can therefore be interpreted as each of the mirrors getting a momentum shift from the photon, independent of the photon's interaction with the other mirrors. This is highlighted in Fig~(\ref{fig:okst}). It is important to note that the momentum imparted to each mirror in a sequence of weak measurements will differ, see Fig~(\ref{see}). This is because the quantum state of the particle changes after interacting with each mirror, due to entanglement with the first mirror. So even with the same photon, the mirror shifts can be slightly different. Though we have ignored correlations, similar to prior investigations \cite{matz1}, some work \cite{Asking-Photons-Where-They-Have-Been, Past-of-a-quantum-particle, Tracing-the-past-quantum-particle, wv24} has considered the slight entanglement between mirrors. 

In the conventional sense, a photon cannot be in two places at once. But it can be in two places in the sense that the photon was where it left a trace. The trace left is slightly entangled such that if robust evidence of this trace is found from an orthogonal component, we will not find the robust evidence in any other place. If the orthogonal component is not detected, the photon's trace can be found in the three distinct locations (A, B, and C) suggesting a simultaneous presence.  So essentially, the entanglement can be seen as $A\perp B C + B\perp A C +C\perp A B$, where we do not consider the E and F terms because they appear as $\epsilon^2$ terms as explained in the preceding section. We conclude that even though the two formalisms are based on different assumptions, they give us consistent results in the first order.

The advantage of using mirrors as probes is that they can help reduce the discrepancies between theory and experiment to some extent. Initially, the mirrors will experience shifts independently of each other. If we have sufficient photons to make a significant momentum transfer, we might observe the mirrors moving slightly. In such a scenario, even if path C were the only sensible path that the photon could have taken, mirrors E and F would still move, regardless of any considered leakages \cite{Past-of-a-quantum-particle}. This would necessitate an optomechanical treatment \cite{PhysRevA.87.043832, new1} to fully understand the implications, including potentially quantum thermodynamic ones \cite{wm17pra}.

\section{Discussion and Conclusion}

The weak value expression we arrived at in Section~(\ref{Particle Paths in a Nested Mach-Zender Interferometer}), can be explained without directly associating particle presence with weak values. We could in principle have a non-vanishing wave function at regions E and  F while also having vanishing weak values \cite{matz1}. This suggests that, from a physical perspective, the particle could be present at E and F, yet detected by a detector different from the one associated with our post-selected state. The slight deviation from the original interpretation of discontinuous trajectories \cite{Asking-Photons-Where-They-Have-Been} is in the sense that the weak trace becomes zero as a result of the pre-selected and post selected wavefunctions being orthogonal at mirrors E and F rather than destructive interference which makes the entire wavefunction disappear. To get a more reliable past of a particle, we could look at all the possible observables associated with the particle along with the spatial projection operator \cite{null, pra24}.

In Section (\ref{Feynman Propagators and Discontinuous Trajectories}), we have shown that the exclusion of the $\epsilon^2$ terms suggests a possibility of the signal at either mirror E or mirror F being too small to be detected. The weak values we have derived are proportional to the weak trace, indicating the presence of particles whenever these values are non-zero \cite{Past-of-a-quantum-particle}. This leads us to suggest again that weak traces of order $\epsilon^2$ could be present, but practically too tiny to be detected against background noise, thus requiring a super-sensitive detector. The `secondary presence' of a particle was proposed in \cite{Tracing-the-past-quantum-particle} where the particle is present in the regions leading to the overlap of the pre-selected and post-selected wavefunctions, however, these regions of secondary presence are too weak to produce a weak trace. If we want to stick to the idea of a photon following discontinuous trajectories, we could in principle attribute the second order terms as the secondary presence. Lastly, the weak value expression derived in Section~(\ref{Particle Paths in a Nested Mach-Zender Interferometer}) takes into account all possible paths inside the interferometer branches, which includes the classical and quantum paths. In more experimentally realisable cases, like the one where we treat the mirrors as the probe, we have a relatively large mass. In such scenarios, the action becomes much larger than 
$\hbar$ and we can modify our equations by adapting semi-classical limit \cite{matz1}. We end by noting that while path integrals allow us to propagate information along continuous paths, it might not be the best approach taken when looking at an interferometer where possible paths traversed by a particle are finite. This was also touched upon by Vaidman \cite{article} in his short response to \cite{Sokolovski} suggesting that any path from the source to the detector is a Feynman path but in the interferometer model, we are not allowed to take paths outside the arms \cite{vaidmanPC, lying24}.

To conclude, we have extended Matzkin's weak value formalism to include subsequent weak measurements in a nested Mach-Zender Interferometer. Our weak value expressions can be interpreted in two different ways. When a photon's constituent property serves as the probe, the resultant weak values aggregate \cite{dada23}, reflecting interference.  On the other hand, treating the mirrors as probes enables the simultaneous independent measurement of mirror states to reflect the weak trace. We further show that non-zero weak traces are based on first-order interactions, and the exclusion of second-order and higher terms, giving vanishing weak values, introduces ambiguity in correlating particle presence with weak values. While theoretically, we could ascribe these higher terms to secondary presence, the practical analysis of their experimental signature would require further study.

\section{Acknowledgements}
We would like to thank Farhan T. Chowdhury for enlightening discussions and helpful suggestions, and for critically reviewing an early draft of the manuscript. We would like to sincerely thank Dr. Alex Matzkin, with whom we had the opportunity to communicate extensively via email, and whose input was invaluable in bringing this project to completion. Last, but not least, we would like to thank Professor Lev Vaidman for a very helpful extended discussion. 

\bibliographystyle{unsrt}
\bibliography{p.bib}

\begin{thebibliography}{10}

\bibitem{ABL}
Yakir Aharonov, David~Z. Albert, and Lev Vaidman.
\newblock How the result of a measurement of a component of the spin of a spin-1/2 particle can turn out to be 100.
\newblock {\em Phys. Rev. Lett.}, 60:1351--1354, Apr 1988.

\bibitem{sup19}
Michael Berry et~al.
\newblock Roadmap on superoscillations.
\newblock {\em Journal of Optics}, 21(5):053002, 2019.

\bibitem{wv21}
Yago~P Porto-Silva and Marcos~C de~Oliveira.
\newblock Theory of neutrino detection: flavor oscillations and weak values.
\newblock {\em The European Physical Journal C}, 81(4):1--10, 2021.

\bibitem{wm21}
Jonathan~T. Monroe, Nicole Yunger~Halpern, Taeho Lee, and Kater~W. Murch.
\newblock Weak measurement of a superconducting qubit reconciles incompatible operators.
\newblock {\em Phys. Rev. Lett.}, 126:100403, Mar 2021.

\bibitem{wv18}
Yakir Aharonov, Eliahu Cohen, Avishy Carmi, and Avshalom~C Elitzur.
\newblock Extraordinary interactions between light and matter determined by anomalous weak values.
\newblock {\em Proceedings of the Royal Society A: Mathematical, Physical and Engineering Sciences}, 474(2215):20180030, 2018.

\bibitem{wmNat19}
KS~Cujia, Jens~M Boss, Konstantin Herb, Jonathan Zopes, and Christian~L Degen.
\newblock Tracking the precession of single nuclear spins by weak measurements.
\newblock {\em Nature}, 571(7764):230--233, 2019.

\bibitem{qu2020sub}
Weizhi Qu, Shenchao Jin, Jian Sun, Liang Jiang, Jianming Wen, and Yanhong Xiao.
\newblock Sub-hertz resonance by weak measurement.
\newblock {\em Nature communications}, 11(1):1752, 2020.

\bibitem{goos47}
Fritz Goos and Hilda H{\"a}nchen.
\newblock Ein neuer und fundamentaler versuch zur totalreflexion.
\newblock {\em Annalen der Physik}, 436(7-8):333--346, 1947.

\bibitem{jorg12}
Mark~R Dennis and Jörg~B Götte.
\newblock The analogy between optical beam shifts and quantum weak measurements.
\newblock {\em New Journal of Physics}, 14(7):073013, jul 2012.

\bibitem{goos13}
J\"{o}rg~B. G\"{o}tte and Mark~R. Dennis.
\newblock Limits to superweak amplification of beam shifts.
\newblock {\em Opt. Lett.}, 38(13):2295--2297, Jul 2013.

\bibitem{wm-goos20}
Akash Das and Manik Pradhan.
\newblock Quantum weak measurement of goos--h{\"a}nchen effect of light in total internal reflection using a gaussian-mode laser beam.
\newblock {\em Laser Physics Letters}, 17(6):066001, 2020.

\bibitem{wm04}
Greg~A. Smith, Souma Chaudhury, Andrew Silberfarb, Ivan~H. Deutsch, and Poul~S. Jessen.
\newblock Continuous weak measurement and nonlinear dynamics in a cold spin ensemble.
\newblock {\em Phys. Rev. Lett.}, 93:163602, Oct 2004.

\bibitem{praQC}
Agata~M. Bra\ifmmode~\acute{n}\else \'{n}\fi{}czyk, Paulo E. M.~F. Mendon\ifmmode~\mbox{\c{c}}\else \c{c}\fi{}a, Alexei Gilchrist, Andrew~C. Doherty, and Stephen~D. Bartlett.
\newblock Quantum control of a single qubit.
\newblock {\em Phys. Rev. A}, 75:012329, Jan 2007.

\bibitem{coto2017power}
Raul Coto, V{\'\i}ctor Montenegro, Vitalie Eremeev, Douglas Mundarain, and Miguel Orszag.
\newblock The power of a control qubit in weak measurements.
\newblock {\em Scientific reports}, 7(1):6351, 2017.

\bibitem{prlQC}
Ping Wang, Chong Chen, Xinhua Peng, J\"org Wrachtrup, and Ren-Bao Liu.
\newblock Characterization of arbitrary-order correlations in quantum baths by weak measurement.
\newblock {\em Phys. Rev. Lett.}, 123:050603, Aug 2019.

\bibitem{wmQC20}
Parveen Kumar, Kyrylo Snizhko, and Yuval Gefen.
\newblock Engineering two-qubit mixed states with weak measurements.
\newblock {\em Phys. Rev. Res.}, 2:042014, Oct 2020.

\bibitem{prxQC}
Farhan~T. Chowdhury, Matt~C.J. Denton, Daniel~C. Bonser, and Daniel~R. Kattnig.
\newblock Quantum control of radical-pair dynamics beyond time-local optimization.
\newblock {\em PRX Quantum}, 5:020303, Apr 2024.

\bibitem{wmQC24}
Philippe Lewalle, Yipei Zhang, and K.~Birgitta Whaley.
\newblock Optimal zeno dragging for quantum control: A shortcut to zeno with action-based scheduling optimization.
\newblock {\em PRX Quantum}, 5:020366, Jun 2024.

\bibitem{Fey48}
Richard~P. Feynman.
\newblock Space-time approach to non-relativistic quantum mechanics.
\newblock {\em Review of Modern Physics}, 20(2s):367--387, 1948.

\bibitem{matz1}
A.~Matzkin.
\newblock Weak values from path integrals.
\newblock {\em Physical Review Research}, 2(3), aug 2020.

\bibitem{Weak-Measurements-Semiclassical-Regime}
A.~Matzkin.
\newblock Observing trajectories with weak measurements in quantum systems in the semiclassical regime.
\newblock {\em Phys. Rev. Lett.}, 109:150407, Oct 2012.

\bibitem{Asking-Photons-Where-They-Have-Been}
A.~Danan, D.~Farfurnik, S.~Bar-Ad, and L.~Vaidman.
\newblock Asking photons where they have been.
\newblock {\em Phys. Rev. Lett.}, 111:240402, Dec 2013.

\bibitem{vonNeumann1955}
John von Neumann.
\newblock {\em Mathematical Foundations of Quantum Mechanics}.
\newblock Princeton University Press, Princeton, NJ, 1955.

\bibitem{col}
Justin Dressel, Mehul Malik, Filippo~M. Miatto, Andrew~N. Jordan, and Robert~W. Boyd.
\newblock Colloquium: Understanding quantum weak values: Basics and applications.
\newblock {\em Rev. Mod. Phys.}, 86:307--316, Mar 2014.

\bibitem{Past-of-a-quantum-particle}
Lev Vaidman.
\newblock Past of a quantum particle.
\newblock {\em Physical Review A}, 87, 04 2013.

\bibitem{Tracing-the-past-quantum-particle}
Lev Vaidman.
\newblock Tracing the past of a quantum particle.
\newblock {\em Physical Review A}, 89, 02 2014.

\bibitem{Time-Symmetry}
Yakir Aharonov, Peter~G. Bergmann, and Joel~L. Lebowitz.
\newblock Time symmetry in the quantum process of measurement.
\newblock {\em Phys. Rev.}, 134:B1410--B1416, Jun 1964.

\bibitem{feynman1965flp}
R.P. Feynman, R.B. Leighton, M.~Sands, and EM~Hafner.
\newblock {\em {The Feynman Lectures on Physics; Vol. I}}, volume~33.
\newblock AAPT, 1965.

\bibitem{book-1}
Richard~P. Feynman and Albert~R. Hibbs.
\newblock {\em Quantum Mechanics and Path Integrals}, chapter 2,6.
\newblock McGraw-Hill, 1965.

\bibitem{book-2}
L.~S. Schulman.
\newblock {\em Techniques and Applications of Path Integration}.
\newblock Dover, Oct,2012.

\bibitem{article1}
Pablo Saldanha.
\newblock Interpreting a nested mach-zehnder interferometer with classical optics.
\newblock {\em Physical Review A}, 89, 02 2014.

\bibitem{Sokolovski}
D.~Sokolovski.
\newblock Asking photons where they have been in plain language.
\newblock {\em Physics Letters A}, 381, 07 2016.

\bibitem{unknown}
Surya~Narayan Sahoo, Sanchari Chakraborti, Som Kanjilal, Saumya~Ranjan Behera, Dipankar Home, Alex Matzkin, and Urbasi Sinha.
\newblock Unambiguous joint detection of spatially separated properties of a single photon in the two arms of an interferometer.
\newblock {\em Communications Physics}, 6(1):203, 2023.

\bibitem{cheshire-cat}
Yakir Aharonov, Sandu Popescu, Daniel Rohrlich, and Paul Skrzypczyk.
\newblock Quantum cheshire cats.
\newblock {\em New Journal of Physics}, 15(11):113015, nov 2013.

\bibitem{classical-limit-of-quantum-optics}
Yakir Aharonov, Alonso Botero, Shmuel Nussinov, Sandu Popescu, Jeff Tollaksen, and Lev Vaidman.
\newblock The classical limit of quantum optics: not what it seems at first sight.
\newblock {\em New Journal of Physics}, 15(9):093006, sep 2013.

\bibitem{opto}
Markus Aspelmeyer, Tobias Kippenberg, and Florian Marquardt.
\newblock Cavity optomechanics.
\newblock {\em Reviews of Modern Physics}, 86, 03 2013.

\bibitem{wv24}
Itamar Stern, Yakov Bloch, Einav Grynszpan, Merav Kahn, Yakir Aharonov, Justin Dressel, Eliahu Cohen, and John~C. Howell.
\newblock Light that appears to come from a source that does not exist.
\newblock {\em Phys. Rev. A}, 109:012206, Jan 2024.

\bibitem{PhysRevA.87.043832}
Chad~R. Galley, Ryan~O. Behunin, and B.~L. Hu.
\newblock Oscillator-field model of moving mirrors in quantum optomechanics.
\newblock {\em Phys. Rev. A}, 87:043832, Apr 2013.

\bibitem{new1}
Raul Corrêa and Pablo Saldanha.
\newblock Photon reflection by a quantum mirror: A wave-function approach.
\newblock {\em Physical Review A}, 93:023803, 02 2016.

\bibitem{wm17pra}
P~Solinas, HJD Miller, and J~Anders.
\newblock Measurement-dependent corrections to work distributions arising from quantum coherences.
\newblock {\em Physical Review A}, 96(5):052115, 2017.

\bibitem{null}
Q.~Duprey and A.~Matzkin.
\newblock Null weak values and the past of a quantum particle.
\newblock {\em Physical Review A}, 95, 11 2016.

\bibitem{pra24}
Jiayu He and Shuangshuang Fu.
\newblock Nonclassicality of the kirkwood-dirac quasiprobability distribution via quantum modification terms.
\newblock {\em Phys. Rev. A}, 109:012215, Jan 2024.

\bibitem{article}
Lü-Bi Deng.
\newblock Theory of atom optics: Feynman’s path integral approach.
\newblock {\em Frontiers of Physics in China}, 1:47--53, 01 2006.

\bibitem{vaidmanPC}
Lev Vaidman.
\newblock Private Communication, Feb. 2024.

\bibitem{lying24}
Lev Vaidman.
\newblock Lying particles.
\newblock {\em Frontiers in Quantum Science and Technology}, 3:1362235, 2024.

\bibitem{dada23}
Jing-Hui Huang, Jeff~S Lundeen, Kyle~M Jordan, Adetunmise~C Dada, Guang-Jun Wang, and Xiang-Yun Hu.
\newblock Weak-value-amplification enhancement of the magneto-optical kerr effect in nanoscale layered structures.
\newblock {\em Physical Review A}, 108(3):033724, 2023.

\end{thebibliography}

\clearpage

\onecolumngrid 
\hoffset = 21pt
\voffset = 17pt
\textheight=649pt
\textwidth= 465pt
\fontsize{10}{14}\selectfont 

\setcounter{figure}{0} \newpage
\setcounter{equation}{0}
\setcounter{section}{0}
\renewcommand{\theequation}{S\arabic{equation}}
\renewcommand{\thefigure}{S\arabic{figure}}

\par{\centering
		{{\large \textbf{Supplementary Material}\\ \bigskip\par\vspace{-0.8em}\large{\underline{Can path integrals justify discontinuous paths in a nested weak value interferometer?}} } \\ \bigskip\par\vspace{-0.8em}Shushmi Chowdhury and Jörg Götte \\ \bigskip\par\vspace{-1.1em}  School of Physics and Astronomy\\ \bigskip\par\vspace{-1.2em} University of Glasgow, Glasgow, G12 8QQ, United Kingdom}
	\bigskip\par}
 \vspace{-2.5em}

\section{Non-separable Propagators and Perturbation}
\label{sec0}
The full propagator, including interaction, is given by
\begin{equation*}
K_{\text{int}}(X_2, x_2, t_2; X_1, x_1, t_1) = \int_{(X_1, x_1)}^{(X_2, x_2)} D[X(t)]D[x(t)] \exp \left[ \frac{i}{\hbar} \int_{t_1}^{t_2} L(X, \dot{X}, x, \dot{x}, t') dt' \right],
\end{equation*}
\begin{equation*}
K_{\text{int}}(X_2, x_2, t_2; X_1, x_1, t_1) = \int_{(X_1, x_1)}^{(X_2, x_2)} D[X(t)]D[x(t)] \exp \left[ \frac{i}{\hbar} \int_{t_1}^{t_2} L_{s} + L_{p} - g(t)A(x)f(x,X_{int})M\dot{X} dt' \right],
\end{equation*}
\begin{equation*}
K_{\text{int}}(X_2, x_2, t_2; X_1, x_1, t_1) = \int_{(X_1, x_1)}^{(X_2, x_2)} D[X(t)]D[x(t)] \exp \left[ \frac{i}{\hbar} \int_{t_1}^{t_2} L_{s} + L_{p} dt' \right] \exp \left[ \frac{-i}{\hbar} \int_{t_1}^{t_2} L_{int} dt' \right],
\end{equation*}

We take the expansion of $L_{int}$ given that the interaction is weak, giving us:
\begin{equation*}
\boxed{
\begin{aligned}
\text{Given that $L_{couple}$ is small, taking first order perturbation}\\
\exp \left[ \frac{-i}{\hbar} \int_{t_1}^{t_2} L_{int} dt' \right] = 1 -  \left[ \frac{i}{\hbar} \int_{t_1}^{t_2}  g(t')A(x)f(x,X_{int})M\dot{X} dt' \right]
\end{aligned}
}
\end{equation*}

Substituting the expansion back in the propagator gives
\begin{align*}
K_{\text{int}} &= \int_{X_1, x_1}^{X_2, x_2} D[X(t)] D[x(t)] \exp\left[\frac{i}{\hbar} \int_{t_1}^{t_2}(L_s + L_p) dt'\right] \times\left[ 1 -  \left[ \frac{i}{\hbar} \int_{t_1}^{t_2}  g(t')A(x)f(x,X_{\text{int}})M\dot{X} dt' \right]\right] \\
&=  \int_{(X_1, x_1)}^{(X_2, x_2)} D[X(t)]D[x(t)] \exp \left[ \frac{i}{\hbar} \int_{t_1}^{t_2} L_s + L_p dt' \right]  \\
&\quad -\frac{i}{\hbar}\int_{t_1}^{t_2}dt'' \left[g(t'')A(x)f(x-X_{\text{int}})M\dot{X}\right] \int_{X_1,x_1}^{X_2,x_2} D[X(t)]D[x(t)] \exp \left[ \frac{i}{\hbar} \int_{t_1}^{t_2} L_{s} + L_{p} dt' \right] \\
&= K_p(X_2,t_2;X_1,t_1)K_s(x_2,t_2;x_1,t_1) \\
&\quad -\frac{i}{\hbar}\int_{t_1}^{t_2}dt''\left[g(t'')A(x)f(x-X_{\text{int}})M\dot{X}\right] \int_{X_1,x_1}^{X_2,x_2} D[X(t)]D[x(t)] \exp \left[ \frac{i}{\hbar} \int_{t_1}^{t_2} L_{s} + L_{p} dt' \right]
\end{align*}

Also we must point out here that the time-dependent interaction constant, g(t''), can be written as $\int_{t_1}^{t_2}g(t'')dt'' = g$, simplifying the equation further to:
\begin{equation}
K_{int} =  K_pK_s  -[\frac{ig}{\hbar}A(x)f(x-X_{\text{int}})M\dot{X} ]\int_{X_1,x_1}^{X_2,x_2} D[X(t)]D[x(t)] \exp [ \frac{i}{\hbar} \int_{t_1}^{t_2} L_{s} + L_{p} dt']
\end{equation}

\subsection{Evolution of the System}
We change the order of integration, following \cite{book-1}, between the path $(t_i,x_i)$ and position variable, $x$.
\begin{equation*}
K^1(b,a) = -\frac{i}{\hbar}\int_{t_a}^{t_b} F(t'') \, dt'', \\
\end{equation*}
where,
\begin{equation}
F(t_c) = \int_{a}^{b} D[q(t)] \exp[\frac{i}{\hbar}\int_{t_a}^{t_b}L_{s}\,dt'] g(t'')A(c)f(x-X_{int})M\dot{X}
\label{s0}
\end{equation}
Changing the order of integration makes it further generalised as,
\begin{equation}
F(t_c) = \int_{-\infty}^{\infty} K_s(b,c)V(c)K_s(c,a) \, dx_c, \\
\end{equation}
Adopting this form to our case, Eq.~{\ref{s0}} gives
\begin{equation}
F(t_c) = \int_{-\infty}^{\infty}  \underbrace{\exp[\frac{i}{\hbar}\int_{t_a}^{t_b}L_{s}\,dt']}_{K_s(b,c)K_s(b,a)} \underbrace{g(t'')A(c)f(q-X_{int})M\dot{X}}_{V(c)} \underbrace{dx
}_{dx_c} 
\end{equation}
\text{We substitute $F(t_c)$ back in $K^1$ which gives the probability amplitude as an ordinary integral }:\\
\begin{equation}
K^1(b,a) =  -\frac{i}{\hbar}\underbrace{\int_{t_a}^{t_b} g(t'')dt''}_{g}\int_{-\infty}^{\infty}  K_s(b,c)A(x_c)f(x-X_{c})M\dot{X}K_s(b,a) dx_c 
\end{equation}
In our model, we have an initial position, the source, and a final position, the detector, so we need not span infinity for integration limits.  Also, $x_c$ and $X_c$ in our model represent the interactions labeled as $\textbf{x}$ and $\textbf{x}$ respectively. Now, putting everything together in $K_v = K^0 + K^1$ gives
\begin{equation}
K_{int} = K_pK_s + \int_{X_1}^{X_2} D[X(t)]e^{{\frac{i}{\hbar}\int L_{p}dt'}} {-\frac{igM\dot{X}}{\hbar}\int_{x_1}^{x_2} K_s(x_2,t_2;x_{c},t_{c}){A(x)f(x,X_{c})}K_s(x_{c},t_{c};x_1,t_1)dx}
\end{equation}
As in \cite{matz1}, $\dot{X}$ at $t_{int}$, is evaluated as $\frac{X(\textbf{t}+\epsilon)-X(\textbf{t})}{\epsilon}$ for each path, with $\epsilon$ being an infinitesimal time variation at $\textbf{t}$ and $M$ is the probe mass.

\section{Subsequent Measurements}
\label{disc}

\begin{figure}[h]
\centering
\includegraphics[width=0.8\linewidth]{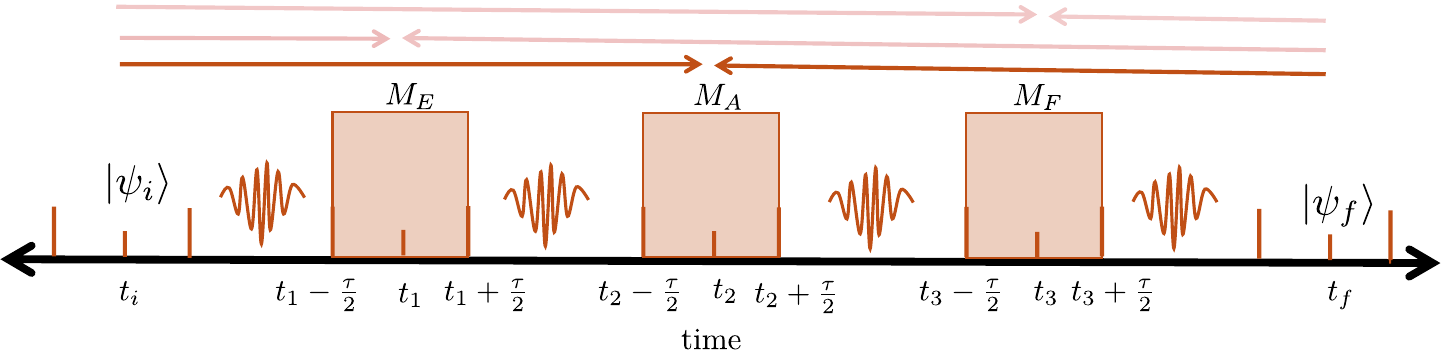}
\vspace{-0.5em}
\caption{Photon scattering through different potentials.}
\vspace{-1em}
\label{fig:sim_mir}
\end{figure}
Fig~(\ref{fig:sim_mir}) outlines the approach when considering mirrors, building on the approach in \cite{Weak-Measurements-Semiclassical-Regime}. The initial state is propagated forward to mirror $M_E$, where the photon, treated as a Gaussian centered at $x_0$, receives a small shift. The final state is propagated back from the detector to mirror $M_F$, where the photon encounters another shift. Subsequently, the shifted probe state from $M_E$ is propagated forward to mirror $M_A$, and similarly, the shifted probe state from $M_F$ is propagated backward to mirror $M_A$. At this mirror, $M_A$, the two shifted wavefunctions superimpose and receive an additional shift from mirror $M_A$. We assume that all interactions at this point occur instantaneously and simultaneously. In the following expressions, $\Omega= \int \frac{L_{interaction}}{\hbar}$ .

\begin{align*}
\Ket{\Psi(t_1 + \tau/2)} &= U(t_1 + \tau/2, t_1) \exp(-i \Omega_1) U(t_1, t_i) \Ket{\Psi(t_i)} \Ket{\phi_1(t_i)} \Ket{\phi_2(t_i)} \Ket{\phi_3(t_i)} \\ \\
\Ket{\Psi(t_2 - \tau/2)} &= U(t_2 - \tau/2, t_1 + \tau/2) \Ket{\Psi(t_1 + \tau/2)} \\ \\
&= U(t_2 - \tau/2, t_1) U(t_1 + \tau/2, t_1) \exp(-i \Omega_1) U(t_1, t_i) \Ket{\Psi(t_i)} \Ket{\phi_1(t_i)} \Ket{\phi_2(t_i)} \Ket{\phi_3(t_i)} \\ \\
&= U(t_2 - \tau/2, t_i) \exp(-i \Omega_1) U(t_1, t_i) \Ket{\Psi(t_i)} \Ket{\phi_1(t_i)} \Ket{\phi_2(t_i)} \Ket{\phi_3(t_i)} \\ \\
\Ket{\Psi(t_2 + \tau/2)} &= U(t_2 + \tau/2, t_2) \exp(-i \Omega_2) U(t_2, t_2 - \tau/2) \Ket{\Psi(t_2 - \tau/2)} \\ \\
&= U(t_2 + \tau/2, t_2) \exp(-i \Omega_2) U(t_2, t_2 - \tau/2) \\ \\
&\quad U(t_2 - \tau/2, t_i) \exp(-i \Omega_1) U(t_1, t_i) \Ket{\Psi(t_i)} \Ket{\phi_1(t_i)} \Ket{\phi_2(t_i)} \Ket{\phi_3(t_i)}
\end{align*}
\begin{align*}
\Ket{\Psi(t_2 + \tau/2)} &= U(t_2 + \tau/2, t_2) \exp(-i \Omega_2) U(t_2, t_1) \exp(-i \Omega_1) U(t_1, t_i) \Ket{\Psi(t_i)} \Ket{\phi_1(t_i)} \Ket{\phi_2(t_i)} \Ket{\phi_3(t_i)} \\ \\
\Ket{\Psi(t_3 - \tau/2)} &= U(t_3 - \tau/2, t_2 + \tau/2) \Ket{\Psi(t_2 + \tau/2)} \\ \\
&= U(t_3 - \tau/2, t_2 + \tau/2) U(t_2 + \tau/2, t_2) \exp(-i \Omega_2) U(t_2, t_1) \exp(-i \Omega_1) U(t_1, t_i) \\ \\
&\quad \Ket{\Psi(t_i)} \Ket{\phi_1(t_i)} \Ket{\phi_2(t_i)} \Ket{\phi_3(t_i)} \\ \\
\Ket{\Psi(t_3 + \tau/2)} &= U(t_3 + \tau/2, t_3) \exp(i \Omega_3) U(t_3, t_3 - \tau/2) \Ket{\Psi(t_3 - \tau/2)} \\ \\
&= U(t_3 + \tau/2, t_3) \exp(-i \Omega_3) U(t_3, t_3 - \tau/2) 
\quad U(t_3 - \tau/2, t_2) \exp(-i \Omega_2)  \\ \\
&\quad U(t_2, t_1) \exp(-i \Omega_1) U(t_1, t_i)\Ket{\Psi(t_i)} \Ket{\phi_1(t_i)} \Ket{\phi_2(t_i)} \Ket{\phi_3(t_i)} \\ \\
\quad \Ket{\Psi(t_3 + \tau/2)} &= U(t_3, t_2) \exp(-i \Omega_3) U(t_3, t_2) \exp(-i \Omega_2) U(t_2, t_1) \\ \\
&\quad \exp(-i \Omega_1) U(t_1, t_i) \Ket{\Psi(t_i)} \Ket{\phi_1(t_i)} \Ket{\phi_2(t_i)} \Ket{\phi_3(t_i)}
\end{align*}

\section{Weak Values for subsequent Measurements }
\label{WS}
We define our initial combined system-probe state as
\begin{equation*}
    \Psi(t_i) = \ket{\psi(t_i)}\ket{\phi_A(t_i)}\ket{\phi_E(t_i)}\ket{\phi_F(t_i)}\ket{\phi_B(t_i)}\ket{\phi_{C}(t_i)},
\end{equation*}
and are interested at first in evaluating the photon when it passes through the probes, $\ket{\phi_{E}} \rightarrow \ket{\phi_{A}} \rightarrow \ket{\phi_{F}} $.
\begin{align*}
\textcolor{black}{\bra{\phi_F}\bra{\phi_E}\bra{\phi_C}\bra{\phi_B}\bra{\phi_A}\bra{\psi_f}} &U^\dagger(t_f,t_3) \exp(-i\Omega_3) U^\dagger(t_3,t_2) \textcolor{black}{\exp(-i\Omega_2)} \\
&U(t_2,t_1) \exp(i\Omega_1) U(t_1,t_i) \textcolor{black}{\ket{\psi_i}\ket{\phi_A}\ket{\phi_B}\ket{\phi_C}\ket{\phi_E}\ket{\phi_F}}
\end{align*}

\begin{align*}
\textcolor{black}{\bra{\phi_F}\bra{\phi_E}\bra{\phi_C}\bra{\phi_B}\bra{\phi_A}\bra{\psi_f}} &U^\dagger(t_f,t_3) U^\dagger(t_3,t_2) \textcolor{black}{\exp(-i\Omega_3)\exp(-i\Omega_2) \exp(-i\Omega_1)} \\
&U(t_2,t_1) U(t_1,t_i) \textcolor{black}{\ket{\psi_i}\ket{\phi_A}\ket{\phi_B}\ket{\phi_C}\ket{\phi_E}\ket{\phi_F}}
\end{align*}

Initial state propagated to interaction time, mirror A, $t_1=t_A$
\begin{equation*}
U(t_1,t_i)\Psi(x_i,t_i) = \int dx_i K_s(x_E, t_1; x_i,t_i)\psi(x_i,t_i)\phi_A(X_A,t_1)\phi_B(X_B,t_1)\phi_C(X_C,t_1)\phi_E(X_E,t_1)\phi_F(X_F,t_1)
\end{equation*}

Final post-selected state propagated back to interaction time, mirror F, $t_3=t_F$
\begin{align*}
U^\dagger(t_{f},t_3)\Psi^*(x_f,t_3) &= \int dx dx_f K^*_s(x_f, t_f; x_F, t_3) b^*_f(t_f, t_3) \bigg(\int_{X_A}^{X_F} D[X_A(t)] e^{\frac{i}{\hbar} \int_{t_i}^{t_f} L_{\phi_A} dt}\bigg) \\
&\quad \times \bigg(\int_{X_B}^{X_F} D[X_B(t)] e^{\frac{i}{\hbar} \int_{t_i}^{t_f} L_{\phi_B} dt}\bigg)
\end{align*}

In the following let $\exp(-i\Omega_n)=\mathrm{Int_n}$. We can expand
\begin{equation*}
\textcolor{black}{\Big[U(t_1,t_i)\Psi(x_i,t_i)\mathrm{}(t_1)\Big]} U(t_2,t_1)\mathrm{Int}(t_2)U^\dagger(t_3,t_2) \textcolor{black}{\Big[\mathrm{Int}(t_3)U^\dagger(t_{f},t_3)\Psi^*(x_f,t_3)\Big]}
\end{equation*}

This can be simplified as
\begin{equation*}
\textcolor{black}{\Big[U(t_2,t_i)\Psi(x_i,t_i)\Big]}\mathrm{Int}(t_1)\mathrm{Int}(t_2)\mathrm{Int}(t_3)\textcolor{black}{\Big[U^\dagger(t_{f},t_2)\Psi^*(x_f,t_3)\Big]}
\end{equation*}

The total Lagrangian contains five interactions with the mirrors:

\begin{equation*}
L_{total} = L_s + L_p + L_{\phi_A} + L_{\phi_B} + L_{\phi_C} + L_{\phi_E} + L_{\phi_F} - L_{\text{int}}{\phi_A} - L_{\text{int}}{\phi_B} - L_{\text{int}}{\phi_C} - L_{\text{int}}{\phi_E} - L_{\text{int}}{\phi_F}
\end{equation*}

\begin{align*}
L_{\text{int}}{\phi_A} &= g(t)A(x)f(x,X_{A})M_A\dot{X}_A = \epsilon_A, \\
L_{\text{int}}{\phi_B} &= g(t)A(x)f(x,X_{B})M_B\dot{X}_B = \epsilon_B, \\
L_{\text{int}}{\phi_C} &= g(t)A(x)f(x,X_{C})M_C\dot{X}_C = \epsilon_C, \\
L_{\text{int}}{\phi_E} &= g(t)A(x)f(x,X_{E})M_E\dot{X}_E = \epsilon_E, \\
L_{\text{int}}{\phi_F} &= g(t)A(X)f(x,X_{F})M_F\dot{X}_F = \epsilon_F.
\end{align*}

For path 1, interaction only takes place in mirrors, $\phi_A$, $\phi_E$, $\phi_F$. In the first order, we have
\begin{equation*}
(1- \epsilon_E)(1-\epsilon_A)(1- \epsilon_F) = 1 - \epsilon_E - \epsilon_A - \epsilon_F + \epsilon_A\epsilon_E + \epsilon_A\epsilon_F + \epsilon_E\epsilon_F - \epsilon_A\epsilon_E\epsilon_F
\end{equation*}
Evaluating the projection of the initial state to the final state $\bra{\Psi(x_f,t_3)} U^\dagger(t_{f},t_2)U(t_2,t_i)\ket{\Psi(x_i,t_i)}$
\begin{align*}
&\int dx_A dx_B dx_C dx_F dx_E K^*_s(x_f, t_f; x_A, t_2) \psi^*_f(t_f, t_2) \\
&\quad \times \left(\int_{X_A}^{X_f} D[X_A(t)] e^{\frac{i}{\hbar} \int_{t_2}^{t_f} L_{\phi_A} dt}\right) \\
&\quad \times \left(\int_{X_B}^{X_f} D[X_B(t)] e^{\frac{i}{\hbar} \int_{t_2}^{t_f} L_{\phi_B} dt}\right) \\
&\quad \times \left(\int_{X_C}^{X_f} D[X_C(t)] e^{\frac{i}{\hbar} \int_{t_2}^{t_f} L_{\phi_C} dt}\right) \\
&\quad \times \left(\int_{X_E}^{X_f} D[X_E(t)] e^{\frac{i}{\hbar} \int_{t_2}^{t_f} L_{\phi_E} dt}\right) \\
&\quad \times \left(\int_{X_F}^{X_f} D[X_F(t)] e^{\frac{i}{\hbar} \int_{t_2}^{t_f} L_{\phi_F} dt}\right) \\
&\quad \times \int dx_A dx_E dx_F K_s(x_A, t_2; x_i,t_i)\psi(x_i,t_i)\phi_E(X_A,t_2)\phi_A(X_A,t_2)\phi_F(X_A,t_2) \phi_B(X_i,t_i)\phi_C(X_i,t_i)\\
&\quad \times \frac{\int dx_A dx_E dx_F dx_f K^*_s(x_f, t_f; x_A, t_2) b^*_f(x, t_w) \textcolor{black}{[1- \epsilon_E - \epsilon_A - \epsilon_F]} \psi(x_A, t_2)}{\int dx_A dx_E dx_F dx_f K^*_s(x_f, t_f; x, t_2) b^*_f(x_A, t_2) \psi(x_A, t_2)}.
\end{align*}

In the above the evolution operators apply to the states E, A and F as we are considering path 1 only. We consider that the probes $\phi_E$ and $\phi_F$ meet $\phi_A$ at the interaction and assume they have already interacted with mirror E and F respectively before this. The previous interactions are taken into account by applying the interaction terms which act as a weighting for all the paths.
\end{document}